\documentclass{aa}

\usepackage{times}
\usepackage{graphics}

\def\Msol{\,{\rm M_{\odot}}}

\def\Rsol{\,{\rm R_{\odot}}}
\begin{document}
\thesaurus{08(08.05.2; 08.11.1; 08.09.2 HIP 60350; 10.15.2 NGC 3603)}
\title{Origin and possible  birthplace of the extreme runaway
star HIP 60350}
\author{P. Tenjes\inst{1, 2}
   \and J. Einasto\inst{2}
   \and H. M. Maitzen\inst{3}
   \and H. Zinnecker\inst{4}}
\offprints{P. Tenjes}
 \mail{tenjes@aai.ee}
\institute{Institute of Theoretical Physics, Tartu University, T\"ahe 4,
           Tartu 51010, Estonia
      \and Tartu Observatory, 61602 T\~oravere, Estonia
      \and Institut f\"ur Astronomie der Universit\"at Wien,
      T\"urkenschanzstrasse 17, A-1180 Wien, Austria
      \and Astrophysikalisches Institut Potsdam, An der Sternwarte 16,
      D-14482 Potsdam, Germany}
\date{Received 28 September 2000 / accepted date}
\maketitle

\begin{abstract}

Using the recently determined spatial velocity components of the
extreme runaway star HIP 60350 and a gravitation potential model of
the Galaxy, we integrate the orbit of HIP 60350 back to the plane of
the Galaxy. In this way, a possible location of the formation of the 
star is determined. We estimate the uncertainty of the result due 
to the uncertainties of the gravitational potential model and the
errors in the spatial velocity components. The place of birth
lies (within the errors) near the position of the open cluster 
NGC 3603. However, the ejection event which occured about 20 Myr
ago is in contradiction with the cluster mean age of $3-4$~Myr. We 
suggest that it occured at an earlier phase in sequential star
formation in that region. We discuss also ejection mechanisms. Due 
to the rather high mass of the star (about $5 \Msol$), the most 
probable model is that of dynamical ejection.

\end{abstract}
\section{Introduction}
Recent Hipparcos proper motion measurements for the star HIP 60350,
together with its radial velocity, allowed the determination of its spatial
velocity components $U=+352\ \mathrm{km~s}^{-1}$, $V=+183\
\mathrm{km~s}^{-1}$ and $ W=+130\ \mathrm{km~s}^{-1}$, giving for the
total velocity the value $v=417\ \mathrm{km~s}^{-1}$ (Maitzen et
al. \cite{mait}). These results suggest the star is a rare
extreme runaway star. The origin of such a high velocity is 
not clear, especially given the rather high mass of the star,
$M\simeq 5\ \mathrm{M}_{\odot }$. Thus, we have chosen to study the
possible origin of the velocity: first the location of birth,
and then the mechanisms allowing the acquisition of such a high
velocity.  Maitzen et al. (\cite{mait}) calculated the distance
of HIP 60350 from the Sun, about 3.5 kpc.  Together with its
equatorial coordinates, this gives us galactocentric cylindrical
coordinates of HIP 60350, $R=9.25\ \mathrm{kpc}$, $\theta =
\mathrm{3.3}^{\mathrm{o}}$, $z=3.4\ \mathrm{kpc}$ (zero point of the
azimuthal angle is in the direction to the Sun). Hence, both the velocities and
the coordinates of the star are known.

The star HIP 60350 is of spectral type B4-5V. The high velocity
carries it far from the galactic plane, towards the outer halo.
For this reason, the study of its orbit is interesting in relation
to another problem, namely the origin of young
stars observed in galactic halos. Recent surveys of the Milky Way high
latitude (halo) regions identified several dozen 
young B-type stars (Saffer et al. \cite{saff}, Lindblad et
al. \cite{lind}, Rolleston et al. \cite{roll}). Within
0.7~kpc from the Sun, Hoogerwerf et al. (\cite{hoog}) identified
56 runaway stars. Slightly more uncertain data indicate that 
there are also young stars in the halo of M~31
(Hambly et al. \cite{hamb}, Smoker et al. \cite{smok}).

The origin of these stars remains unclear, as well as their place of
birth. Two main theories exist. According to the first theory these
stars are real halo stars, i.e. they were born at high galactic
latitudes as a result of collisions between high-velocity clouds
(e.g. Dyson \& Hartquist \cite{dy:ha}). However, for several stars of
this kind their radial velocities indicate that they are moving away
from the Galactic plane with high speed. Thus the second theory is
based on the hypothesis that these stars are ejected with high
velocities from the disk by some mechanism. Two main mechanisms have
been considered so far -- ejection via supernova explosion (Blaauw
\cite{blaa}), and dynamical ejection via close stellar encounters in
star clusters (see e.g. Leonard \& Duncan \cite{le:du1}).
Unfortunately, spatial velocities of these stars are poorly known and
even if known they may include systematic errors (Rolleston et
al. \cite{roll}).

For this reason, the study of HIP 60350 provides a good opportunity to
determine the possible place of birth. In an earlier study, Maitzen et
al.  (\cite{mait}), using simple dynamical considerations, estimated
the time needed for the star to move from the galactic plane to its
present position (about 20 Myrs), and speculated that
the possible place of birth could be the spiral arm -II in the 4th
quadrant. In the present study we use a detailed gravitational
potential model of the Milky Way, integrate numerically the velocity
components and coordinates backwards, and determine the place of birth
in the galactic plane. We estimate also the uncertainties of the
result and look for nearby star clusters.
\section{Gravitational potential of the Galaxy}
Numerical calculation of a stellar orbit presumes the knowledge
of the gravitational potential of the galaxy. In the present
paper, our main interest is not the long-term parameters of an orbit,
i.e. the behaviour of the orbit in the far future. This kind of
analysis depends essentially on the mass distribution of the
Galaxy at large galactocentric distances, particularly on the
distribution model of dark matter. Rather, our main aim here is to find
the possible place of birth of the star. For this, it is important
to know the mass distribution of the Milky Way mainly at
intermediate distances.  For that region, an accurate rotation
curve of the Galaxy exists (construction of a mass distribution
model of the Galaxy includes, beside the rotation curve, the
knowledge of additional parameters, e.g. the local Oort constants,
the solar distance and the circular velocity etc., see Einasto
(\cite{e79}) or a more recent review by Dehnen \& Binney
(\cite{de:bi})).  The circular velocity is directly related to the
radial derivative of the gravitational potential, which is
contained in the equations of motion of a star. Several models
exist which describe the gravitational potential of the Galaxy.
In order to calculate the orbit of the star HIP 60350, we
need to know the radial and vertical derivatives of the
potential also outside the galactic plane. Hence, it is better to
use models which take into account the real ellipticity of the
Milky Way mass distribution, including the individual flatness of
different stellar populations. For this reason we decided to use
the model of the Galaxy proposed by Haud \& Einasto
(\cite{ha:ei}).  A recent model by Dehnen \& Binney
(\cite{de:bi}) also takes into account the real flatness of
stellar populations but their algorithm for gravitational
potential calculation is less convenient for our purpose.

In our model, the Galaxy is given as a superposition of different
subsystems. Each subsystem represents a certain stellar/gas population
with corresponding density distribution, chemical composition and
kinematical characteristics. The density distribution of each
component is approximated by an inhomogeneous ellipsoid of rotational
symmetry with constant axial ratio $\epsilon$ (Einasto \cite{eina},
Einasto \& Haud \cite{ei:ha}).  The spatial density of visible
populations is described by the law
$$\rho (a)=\rho (0)\exp [-(a/a_c)^{1/N}], \eqno(1)$$
where $\rho (0)=hM/(4\pi\epsilon a_0^3)$ is the central density, $a=
\sqrt{R^2+z^2/\epsilon^2}$ is the distance along the major axis,
$a_c=ka_0$ is the core radius ($ a_0$ is the harmonic mean radius),
$h$ and $k$ are normalizing parameters, depending on the parameter
$N$, which allows to vary the density behaviour with $a$.  The
definition of normalizing parameters and their calculation is
described in Tenjes et al. (\cite{tenj4}), appendix B. For the disk
and the flat components we use the density distribution in the
following form
$$\rho (a)=\rho _{+}(a)-\rho _{-}(a), \eqno(2)$$
where subindices ``$+$'' and ``$-$'' denote density distributions (1)
of components with positive and negative masses respectively. In this
way, we obtain density distributions with a central density depression.
If we demand that the density be zero at $a=0$ and positive elsewhere,
the following relations must hold between the parameters of components
$\rho_+$ and $\rho_-$: $ a_{0-}=\kappa a_{0+}$, $M_-=-\kappa^2M_+$,
$\epsilon _-=\epsilon_+/\kappa ,$ where $\kappa <1$ is a parameter
which determines the relative size of the hole in the centre of the
disk. To avoid negative densities of the population, the structural
parameters of components $N_+$ and $N_-$ must be equal (see Einasto et
al.  \cite{einaa}).

The dark matter distribution is represented by a modified isothermal
law
$$\rho (a) = \cases{\rho (0) ([1+({a\over a_c})^2]^{-1/2} -
                      [1+({a^0\over a_c})^2]^{-1/2})^2 & $a \leq a^0$
                      \cr
                    0         &      $a>a^0$.    \cr } \eqno(3) $$
Here $a^0$ is the outer cutoff radius of the isothermal sphere.

The modeling procedure and model parameters are given in the original
paper and it is not necessary to repeat them here. Knowing the
density distribution formula of the components and the parameters of the
populations we can calculate the gravitational potential at every
point $(R,z)$.
\section{Numerical orbit calculation}
The equations of motion in cylindrical coordinates are (see e.g. Binney \&
Tremaine \cite{bi:tr}, Sect. 3.1)
$$L_z = Rv_{\psi} = {\rm const},$$
$$\ddot{R} = -{\partial\Phi\over\partial R} + {L^2_z \over R^3},
\eqno(4)$$
$$\ddot{z} = -{\partial\Phi\over\partial z},$$
where $\Phi$ is the gravitational potential not depending on the azimuthal
coordinate $\psi$.  The gravitational potential derivatives for
inhomogeneous ellipsoidal mass distribution with constant ellipticity
are
$${\partial\Phi (R,z) \over \partial R} = R {GhM\over (ea_0)^3}
\int_0^{\arcsin (e)} \rho^* (a) \sin^2 x\ dx \eqno(5a)$$
$${\partial\Phi (R,z) \over \partial z} = z {GhM\over (ea_0)^3}
\int_0^{\arcsin (e)} \rho^* (a) \tan^2 x\ dx \eqno(6a)$$
where $a^2 = {\sin^2 x\over e^2} \left( R^2 + {z^2\over\cos^2 x}
\right)$, $e = \sqrt{1-\epsilon^2}$ is the eccentricity and $\rho^*
(a) = \exp \left( - \left[ a/ (ka_0)\right] ^{1/N} \right) $.
Here $\rho^* (a)$ is  our density distribution (1).

For a spherical dark matter distribution the gravitational potential
derivatives are
$${\partial\Phi (R,z) \over \partial R} = R {GhM\over a_0^3} \int_0^1
\rho^* (a)\ u^2\ du, \eqno(5b)$$
$${\partial\Phi (R,z) \over \partial z} = z {GhM\over a_0^3} \int_0^1
\rho^* (a)\ u^2\ du, \eqno(6b)$$
where $a^2 = u^2 (R^2+z^2)$ and $\rho^* (a)$ is again the density
distribution (3) without the factor $\rho (0)$.

We calculated integrals (5) and (6) numerically using the Gaussian
quadrature formula with 40 points.  The equations of motion (4) were
solved numerically using the 4th order Runge-Kutta method.

The correctness of the calculated orbits was tested in several ways.
The most simple test is to calculate the circular velocity at some
distance (e.g. at solar distance) with the known formula $V^2 = R
\partial\Phi /\partial R$, and then integrate the orbit numerically
with these initial data. Doing so we found that the orbit remains
circular with a precision of 0.1 percent.

\begin{figure}
  \resizebox{\hsize}{!}{\includegraphics{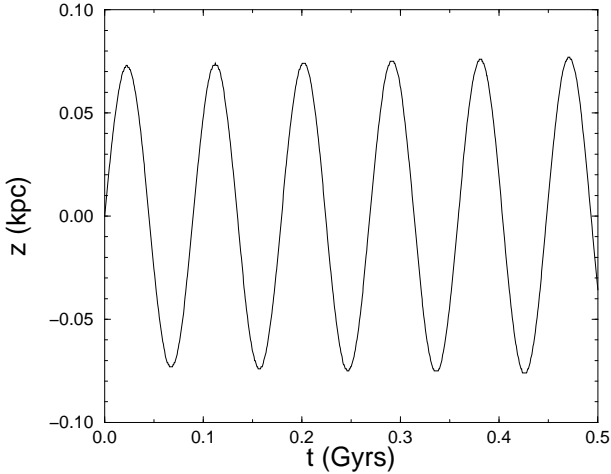}}
  \caption[ ]{Test orbit with $(v_R, v_{\psi}, v_z) = (0, 217, 5)$.}
  \label{figu1}
\end{figure}

Next we calculated vertical oscillations relative to the galactic
plane at solar distances with $v_z =$ $5\ \mathrm{km~s^{-1}}$ at $z=0$.  The
orbit is presented in Fig.~\ref{figu1}.  We see that the period of
vertical oscillations remains constant, as it must be, and is $T =
0.0896$ Gyrs. For small oscillations this period is simply
related to the local mass density:
$$\rho = {\pi \over GT^2} + {A^2-B^2 \over 2\pi G}.$$
Here $A =$ $14.8\ \mathrm{km~s^{-1}kpc^{-1}}$ and $B =$ 
$-12.4\ \mathrm{km~s^{-1}kpc^{-1}}$ are
Oort's constants (Feast \& Whitelock \cite{fe:wh}), giving the value $\rho =$
$0.0890\ \mathrm{\Msol pc^{-3}}$. The direct calculation from the initial
parameters of the galactic model yields as the local mass density
$0.0884\ \mathrm{\Msol pc^{-3}}$, in very good agreement with the
previous value.

\begin{figure}
  \resizebox{\hsize}{!}{\includegraphics{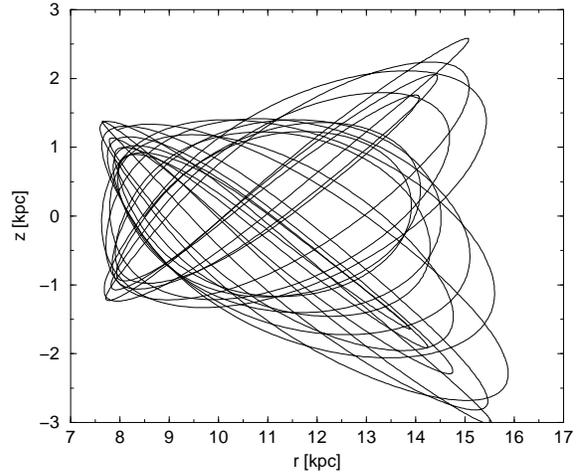}}
  \caption[ ]{A meridional section of a test orbit with $(v_R,
  v_{\psi}, v_z) = (50, 250, 50)$.}
  \label{figu2}
\end{figure}

Finally, we calculated a typical orbit with $(v_R, v_{\psi}, v_z) =
(50, 250, 50)$. This orbit in the comoving meridional plane is
presented in Fig.~\ref{figu2} . We see that the orbit remains well
confined within a region limited by the integrals of motion (see
Kuzmin \cite{kuzm}, Ollongren \cite{ollo}).

After these tests we began to calculate the orbit of HIP 60350.
\section{Orbit of the star HIP 60350}
\begin{figure}
   \resizebox{\hsize}{!}{\includegraphics{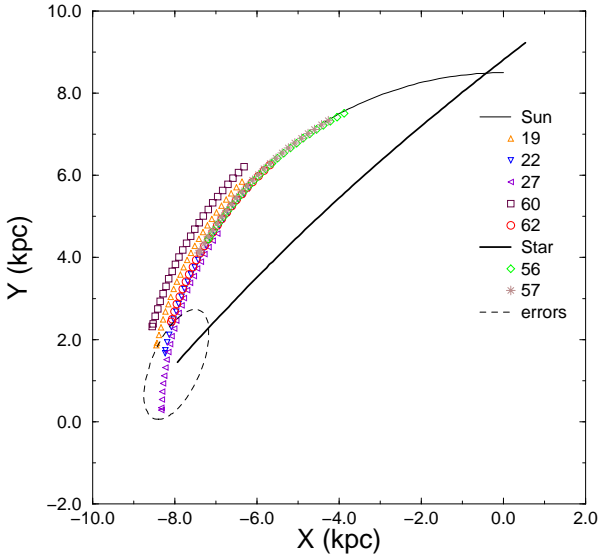}}
    \caption[
   ]{Motion of clouds Georgelin 56, 57, 60 and 62 (=NGC 3603),
   Grabelsky 19, 22, and 27, the Sun and the star HIP 60350 at the
   Galactic plane during 20.4~Myrs. For HIP 60350 the orbit projected
   to the Galactic plane is given.}
\label{figu3}
\end{figure}

Knowing the present phase space coordinates for HIP 60350 we
calculated its orbit back, and found that the star was in the galactic
plane 20.4 Myrs ago at $R = 8.05$~kpc and $\theta = -80^{\circ}$.
The spatial velocity components at that moment were $v_r =$ $-278\ 
\mathrm{km~s}^{-1}$, $v_{\theta} =$ $463\ \mathrm{km~s}^{-1}$, 
$v_z =$ $191\ \mathrm{km~s}^{-1}$. Subtracting the galactic
rotation component at 8~kpc $v_{\theta}^0 =$ $220\ \mathrm{km~s}^{-1}$ 
we obtain that the ejection velocity was $v =$ $416\ \mathrm{km~s}^{-1}$. 
The projection of the orbit into the plane of the Galaxy is shown in 
Figure~\ref{figu3} by a bold solid line.

In order to examine possible ejection mechanisms it is needed to
estimate the uncertainties in our result. First, we estimate
errors due to the galactic mass model. An acceptable model must be in
accordance with measured galactic rotation velocities.  Thus, the mass
of the galactic disk (which is the most essential parameter in our
case) must give rotation velocities within the measured velocity
errors. The rotation curve at distances $R \sim 8-10$~kpc from the
Galactic centre is known with errors $\pm 10\ \mathrm{km~s}^{-1}$ 
(Fish \& Tremaine
\cite{fi:tr}, Binney \& Merrifield \cite{bi:me}). These errors allow
us to vary the disk mass within the limits $\pm 0.9\cdot 10^{10}
\Msol$. Orbit calculations for galactic models with disk masses
$5.9 \cdot 10^{10} \Msol$ and $7.8\cdot 10^{10}\Msol$ show that the
position of galactic plane crossing has an uncertainty of only $\pm
0.1$~kpc, which is rather small.

Second, we studied uncertainties due to the errors in the observed
velocity components of the star, quoted by Maitzen et al.
(\cite{mait}) to be 15 percent. To the $\theta$-component velocity we
must add the circular velocity of LSR ($220\pm 10\ \mathrm{km~s}^{-1}$).  
Thus the uncertainties of the velocity components are $(\pm 50, \pm 30, \pm
20)$. To estimate the influence of velocity uncertainties of each
component on the position of the star at the galactic plane crossing, we
calculated orbits of the star for a number of observed velocities
including estimated errors.  We find that the resulting overall
uncertainty of the plane crossing lies approximately within an ellipse
with semiaxes 1.1 kpc and 0.7 kpc.  The long axis of the ellipse is
oriented approximately in the direction of the stellar orbit
(Fig.~\ref{figu3}, dashed ellipse). Hence we must search for the
possible birthplace of HIP 60350 within this region.

Because our star is young, we are interested in the presence of
young star clusters and associated interstellar gas regions.
According to a catalog of HII regions of the Milky Way by Georgelin
\& Georgelin (\cite{ge:ge}), potential candidates could be objects
Nos. 56, 57, 60 and 62. In addition, we looked for promising
candidates in the molecular gas survey by Grabelsky et al.
(\cite{grab}); clouds Nos. 19, 22 and 27 are
candidates.  From these lists only clouds No. 60 and 62 are
known to contain stars at present. Moreover, cloud No. 62
corresponds to the well-known star cluster NGC~3603; thus, the
distance to this cluster is known with high accuracy. The
distances to other clouds without known associated stars are
estimated by the authors of corresponding papers using 
indirect data and are less certain (distances to the clouds Nos
56 and 57 are estimated by Georgelin \& Georgelin (\cite{ge:ge}) on
the basis of galactic northern and southern rotation models,
distances to the clouds Nos 19, 22 and 27 are estimated by Grabelsky
et al. (\cite{grab}) on the basis of radius--line-width relation; in both
papers the authors used the old values of galactic constants (10 kpc and
$250\ \mathrm{km~s}^{-1}$), thus we transformed the clouds distances to
the new values of galactic constants recommended by the IAU, see Kerr
\& Lynden-Bell (\cite{ke:ly})). As the star HIP 60350 started from the
galactic plane about 20~Myr ago we must transfer the positions of the
selected clouds back by that time assuming circular orbits at
their galactocentric distances. The results are presented in
Fig.~\ref{figu3}.

The positions of young objects in the galactic plane may be
influenced by their peculiar velocity component in the
$R$-direction. Peculiar velocities of young star clusters are
typically $\le 15\ \mathrm{km~s}^{-1}$ (see e.g. Nezhinskij et al.
\cite{nezh}), giving during a 20~Myr orbit a maximum correction
of the position up to $\pm 0.3$~kpc. Thus we see that clouds Nos.
60 and 62, which are known to contain stars, as well as
molecular clouds, may lie inside the boundary of the permitted
region of the birthplace of our star.

Recent observations (Brandl et al. \cite{bran}, De Pree et al.
\cite{depr}) allowed us to study in detail the star cluster NGC 3603.
According to these observations this cluster lies at a distance $6.1
\pm 0.6$~kpc from the Sun. NGC 3603 contains a lot of young and
massive stars and is still in the stage of star formation. Its
position corresponds to the birthplace of HIP 60350 nearly perfectly.
Related to the cluster HII region is one of the largest HII regions in
the Galaxy. The initial mass function (IMF) of the cluster extends up
to $120 \Msol$ (Drissen et al. \cite{dris}).
\section{Discussion}
In the present work we integrated the orbit of the extreme runaway
star HIP 60350 back to the galactic plane. Taking into account
possible errors in the velocity measurements and the galactic model we
have found the probable birthplace of this star.

OB runaway stars can be produced either by supernova explosions in
massive close binaries or by close dynamical encounters in dense star
clusters.  First we discuss the ejection as a result of SN explosion.
This mechanism was studied first by Blaauw (1961) as a symmetric SN
explosion where the momentum of ejected matter was balanced by the
momentum of the binary centre-of-mass. However, before the SN
explosion there must occur some mass transfer between the system
components, and thus the amount of ejected matter is less than half of
the system's total mass. Disruption of a binary system is unlikely and
the maximum velocity of the centre-of-mass cannot be too large.
According to Tauris \& Bailes (\cite{ta:ba}), the limit for the recoil
velocity is $270\ \mathrm{km~s}^{-1}$ (see also Nelemans et al. \cite{nele}).

According to recent studies, supernova explosions can be asymmetric.
To explain the observed spatial velocity distribution of single
pulsars, we need to assume asymmetric SN explosions, where the
newborn neutron star, due to an additional kick, attains a mean
velocity of $450\ \mathrm{km~s}^{-1}$ in an arbitrary direction (Lyne \& Lorimer
\cite{ly:lo}, Hartmann \cite{hart}). This explosion probably disrupts
the system, and thus runaway stars should not be presently binary
stars. Although an asymmetric explosion may give a very high velocity
to the pulsar (up to $1500\ \mathrm{km~s}^{-1}$), its impact to the companion star is
significantly smaller.  According to calculations by Tauris \& Takens
(\cite{ta:ta}) the spatial velocity of the companion star depends on
several parameters (kick speed, companion star mass etc); usually it
does not exceed $300\ \mathrm{km~s}^{-1}$. Applying these results to our $5\Msol$ star,
the runaway velocity is less than $200\ \mathrm{km~s}^{-1}$, even for the
largest kick speeds.

Because it is difficult to obtain the ejection velocity of $420\ 
\mathrm{km~s}^{-1}$ with a SNe scenario, we turn our attention to dynamical
interactions. Dynamical interactions have been studied by Leonard \&
Duncan (\cite{le:du1}), who showed that typical velocities in
interactions extend up to $250\ \mathrm{km~s}^{-1}$. According to models the
most efficient in producing runaway stars are binary-binary interactions
(Mikkola \cite{mikk}). The efficiency of binary-binary interactions
dominates over binary-single interactions, especially in producing
high-velocity runaway stars (Leonard \& Duncan \cite{le:du1}). However,
these are the results of N-body experiments and were limited by
available computational time. For this reason Leonard (\cite{leon})
performed special numerical experiments with binary-binary
interactions to determine the maximum possible velocity of ejection.
His results were presented for different ratios $v_{ej}/v_{esc}$
($v_{ej}$ -- ejection velocity of a star, $v_{esc}$ -- escape velocity
at the surface of ejected star).

HIP 60350 was ejected from the galactic plane with a spatial velocity of
$420\ \mathrm{km~s}^{-1}$. A $5 \Msol$ star has a radius of $2.6 \Rsol$ 
(Tout et al. \cite{tout}) and $v_{esc} =$ $840\ \mathrm{km~s}^{-1}$, 
thus $v_{ej} = 0.50 v_{esc}$. According to Leonard (\cite{leon}, his 
Table 3) the ejection velocity $v_{ej} = 0.50 v_{esc}$ cannot be obtained 
if HIP 60350 interacted with equal or lower mass stars. Such high
ejection velocities can be obtained by a star in interactions with
three stars having masses at least 4 times larger, but probably even 8
times larger. In this case sufficient ejection velocities for the
least massive companion may appear within about 400 cluster
crossing times. Hence to produce a high-velocity star like HIP
60350, the participation of 3 stars with masses $\sim 40\Msol$ is
needed. Typical diameters of young open clusters are $1-9$~pc 
(Phelps \& Janes \cite{ph:ja}), the line-of-sight velocity
dispersion is about $3-4\ \mathrm{km~s}^{-1}$ (see e.g. Kroupa \cite{krou}), 
thus the resulting crossing time is $0.3-2$~Myr. On the other hand,
according to photometric measurements, HIP 60350 is clearly rather
close to the ZAMS, and cannot be old (Maitzen et al. \cite{mait}). It
is very plausible that the ejecting event took place in an early phase
of star formation in a very compact cluster, where IMF was skewed
towards the high-mass end compared with the field star IMF (Clarke \& 
Pringle \cite{cl:pr}). If this
is the case, stars of larger mass could indeed participate in the ejection
event (MS lifetimes of $20\Msol$ and $40\Msol$ stars are 5~Myr and 2~Myr,
respectively).

\begin{figure}
   \resizebox{\hsize}{!}{\includegraphics{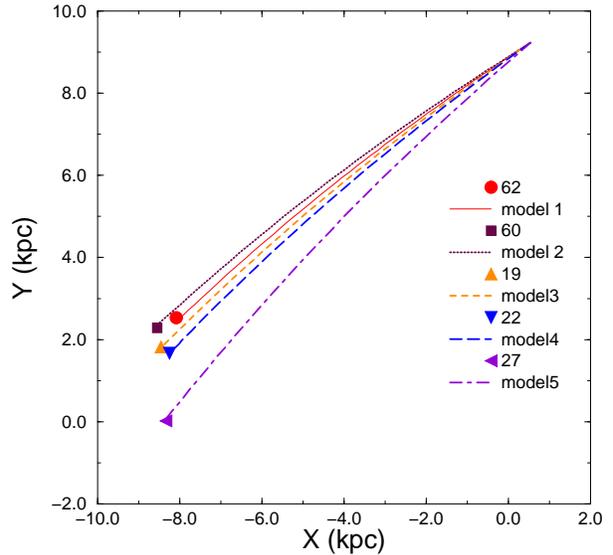}}
   \caption[ ]{Positions of clouds at the times of star ejection.
   Corresponding orbits of the ejected star are labeled as different
   models with parameters given in Table~1.}
   \label{figu4}
\end{figure}

Dynamical interactions are more efficient in massive and compact
clusters (Leonard \& Duncan \cite{le:du2}). Interactions (ejections,
mergers) are also more frequent in clusters with active star 
formation at early evolutionary stages (Bonnell et al. \cite{bonn},
Portegies Zwart et al. \cite{port}). For this reason the cluster G~62 =
NGC~3603 seems to be a suitable candidate. The IMF of the cluster extends
up to large masses, making high ejection velocities reasonable. When
varying the observed velocity of HIP 60350 (but remaining within
observational errors) we calculated that with the present velocity
components $v_r =$ $318\ \mathrm{km~s}^{-1}$, $v_{\theta}=$ $423\ 
\mathrm{km~s}^{-1}$ and $v_z =$ $138\ \mathrm{km~s}^{-1}$ the place 
of birth of the star coincides with the position of G62 
(Fig.~\ref{figu4}, circle and continuous line, see also Table~1,
Model~1).

However, there exists a serious argument against NGC 3603 as the
birthplace of HIP 60350. The age of NGC 3603 is estimated to be
only $3-4$ Myrs (De Pree et al.  \cite{depr}), which is in
contradiction to an ejection 20 Myrs ago. Further, some time
is also needed for interactions -- statistically up to $400
t_{\rm cross} \ge 100$~Myr. This discrepancy may not be critical, 
because star formation in young clusters is a complicated
process with different stages (see e.g. Elmegreen \cite{elme} 
and for NGC 3603, Brandner et al. \cite{brand}). Thus,
we would like not to exclude this cluster together with its
surrounding region as a candidate for the origin of HIP 60350.
The age structure of the cluster deserves special and careful
study: Eisenhauer et al. \cite{eise} derived that the
distribution of stellar ages in NGC 3603 is non-Gaussian and 
extends up to 100~Myr.

\begin{table}
 \caption[]{Parameters of stellar orbits for individual clouds}
 \begin{flushleft}
 \begin{tabular}{llllll}
\hline\noalign{\smallskip}
Model&\multicolumn{3}{c}{Present velocity (km/s)} &Ejection&Ejection\\
\cline{2-4}
     & $v_r$ &$v_{\theta}$&$v_z$& time (Myr)& velocity \\
      \noalign{\smallskip}\hline\noalign{\smallskip}
1 &  318 &  423 &  138 & 19.9 & 405 \\
2 &  307 &  428 &  130 & 20.7 & 402 \\
3 &  332 &  423 &  130 & 20.7 & 416 \\
4 &  343 &  418 &  130 & 20.5 & 420 \\
5 &  397 &  403 &  117 & 21.6 & 456 \\
\noalign{\smallskip}\hline
\end{tabular}
\end{flushleft}
\end{table}

\begin{figure}
   \resizebox{\hsize}{!}{\includegraphics{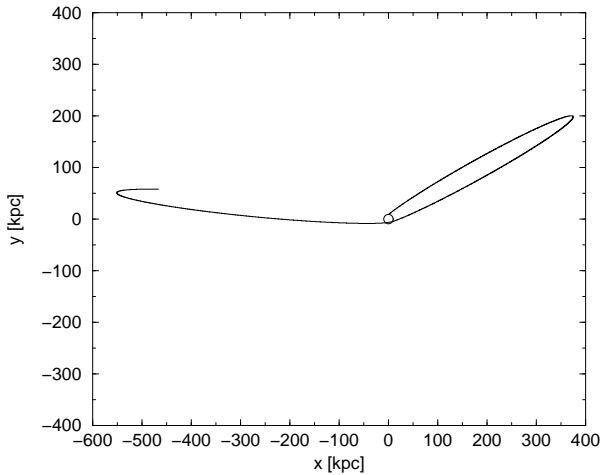}} \caption[ ]{The
   orbit of the star HIP 60350 for the next 13.5~Gyrs, projected to the
   galactic plane. Circle with centre coordinates $(0,0)$ is situated
   at the galactic centre and has a radius 10~kpc.  The $x$ and $y$
   axes are directed toward the direction of rotation (at Sun
   position) and away from the centre toward the Sun, respectively.}
\label{figu5}
\end{figure}

Another candidate is the HII region G60, which is also situated
quite near to the probable ejection place. With present velocity
components $v_r =$ $307\ \mathrm{km~s}^{-1}$, $v_{\theta}=$ $428\ 
\mathrm{km~s}^{-1}$ and $v_z =$ $130\ \mathrm{km~s}^{-1}$ the 
ejection place will lie close to G60
(Fig.~\ref{figu4}, Table~1, model 2). The HII region G60 is less
studied. We do not know the detailed structure and age of this
region. These properties remain suitable for further study.

According to the study of molecular clouds by Grabelsky et al.
(\cite{grab}), near the birthplace of HIP 60350 there are clouds of
molecular gas No 19, 22 and 27 according to his designations (in
addition cloud No 17 corresponds to NGC~3603). For these clouds the
corresponding models are 3 -- 5, see Table~1.  Unfortunately the
distances to these clouds are kinematic (they have based on empirical
radius--line-width relations by Dame et al. (\cite{dame}), see
Grabelsky et al.) and are thus uncertain. Moreover, these
clumps may be blended together at lower intensities, and SN explosions
(\"Opik \cite{opik}) and stellar winds from massive stars (Carpenter
et al. \cite{carp}) may disrupt the original cloud
structure. Molecular clouds have lifetimes of $10-100$~Myr (Blitz \&
Williams \cite{bl:wi}, Williams et al. \cite{will}) or less
(Ballesteros-Paredes et al. \cite{ball}, Elmegreen \cite{elme}), which
is probably too short to produce high ejection velocities. Thus, it is
not likely that the referred molecular clouds can be directly assigned
as birthplaces of HIP 60350. 

At present the place of birth of our star is at $l=298$, $b=0$
and $R=6.9$~kpc from the Sun. This region (see also the
ellipse of errors in Fig.~\ref{figu3}) deserves further careful
study. According to our analysis it is probable that in this
region there is an open cluster with an age of more than 20~Myr.
Finding this kind of star cluster is not simple, due to the
relatively large distance and possible ISM obscuring (unless we
are lucky enough to find it in a cloud hole). In order to decrease the
probable search area from where the ejection took place it seems
most promising to determine more precise velocity measurements
for the star.

Individually identified molecular clouds referred to above may 
simply be remnants of an earlier star formation process.
Together with NGC~3603 they may form a bigger stellar-gaseous
complex. Observations by Grabelsky et al. (\cite{grab}) hint that
near NGC~3603 a more massive and extremely disrupted cloud
complex exists which refers to past intensive star formation. 
Massive stars from this first star formation stage ejected 
HIP~60350, exploded as supernovae and triggered star formation in 
the adjecent parts of the original cloud. These events might be similar 
to the scenario outlined by Preibisch \& Zinnecker (\cite{pr:zi}) for 
Scorpius-Centaurus OB association.

Observations of {\em different individual} OB runaways support both
the supernova explosion scenario (e.g. Kaper et al. \cite{kape}) and
the cluster ejection model (e.g. Ryans et al. \cite{ryan}, Moffat et
al. \cite{moff}); both scenarios produce runaways (see recent
detailed analysis about nearby runaways by Hoogerwerf et al.
\cite{hoog}). The very high velocity runaway HIP 60350 seems to been
ejected via dynamical ejection from G62 or from G60.

Finally, we have calculated the future orbit of the star HIP~60350,
starting from its present position. The total galactocentric velocity
of the star is less than the escape velocity at $R=9$~kpc, $600\
\mathrm{km~s}^{-1}$, thus it must remain within the potential well of the
Galaxy.  The projection of the orbit to the galactic plane is shown in
Fig.~\ref{figu5}. We see that its apogalactic distance reaches in the first and
second revolution 425 and 554 kpc at 3.3 and 11.4 Gyrs, respectively.
Evidently, at these large distances, perturbations by other members of
the Local Group galaxies are important.  Most likely the star will
remain in the common potential well of the whole Local Group.

\begin{acknowledgements}

We would like to thank the anonymous referee for useful comments and
suggestions. PT acknowledges financial support from the Estonian
Science Foundation (grant 2627); JE acknowledges financial support
from the Estonian Science Foundation (grant 2625) and hospitality
during visits to Institut f\"ur Astronomie der Universit\"at Wien.

\end{acknowledgements}


\begin{thebibliography}{}
%
\bibitem[1999]{ball}
Ballesteros-Paredes J., Hartmann L., V\'azquez-Semanini
E., 1999, ApJ 527, 285
%
\bibitem[1998]{bi:me}
Binney J., Merrifield M., 1998, Galactic astronomy, Princeton
Univ. Press, Princeton, New Jersey
%
\bibitem[1987]{bi:tr}
Binney J., Tremaine S., 1987, Galactic dynamics, Princeton Univ.
Press, Princeton, New Jersey
%
\bibitem[1961]{blaa}
Blaauw A., 1961, BAN 15, 265
%
\bibitem[2000]{bl:wi}
Blitz L., Williams J.P., 2000. Molecular clouds. In:
Kylafis N., Lada C.J. (eds.), The physics of star formation and
early stellar evolution. (in press), astro-ph/9903382
%
\bibitem[1998]{bonn}
Bonnell I.A., Bate M.R., Zinnecker H., 1998,
MNRAS 298, 93
%
\bibitem[1999]{bran}
Brandl B., Brandner W., Grebel E.K., Zinnecker
H., 1999, The Messenger, Nr 98, 46
%
\bibitem[1997]{brand}
Brandner W., Grebel E.K., Chu Y-H., Weiss K., 1997, ApJ
475, L45
%
\bibitem[1995]{carp}
Carpenter J.M., Snell R.L., Schloerb F.P., 1995, ApJ 450,
201
%
\bibitem[1992]{cl:pr}
Clarke C.J., Pringle J.E., 1992, MNRAS 255, 423
%
\bibitem[1986]{dame}
Dame T.M., Cohen R.S., Elmegreen B.G., Thaddeus P., 1986,
ApJ 305, 892
%
\bibitem[1998]{de:bi}
Dehnen W., Binney J., 1998, MNRAS 294, 429
%
\bibitem[1999]{depr}
De Pree C.G., Nysewander M.C., Goss W.M., 1999, AJ 117, 2902
%
\bibitem[1995]{dris}
Drissen L., Moffat A.F.J., Walborn N.R., Shara M.M., 1995, AJ
110, 2235
%
\bibitem[1983]{dy:ha}
Dyson J.E., Hartquist T.W., 1983, MNRAS 203, 1233
%
\bibitem[1972]{eina}
Einasto J., 1972. Tartu Astr. Obs. Teated 40, 3; also in: Stars
and the Milky Way System, Proc. First European Ast. Meet. vol 2,
ed. L.N. Mavridis, Springer, Berlin, Heidelberg, New York, p. 291
%
\bibitem[1979]{e79}
Einasto J., 1979. Galactic mass modeling. In: Burton W.B. (ed.),
The large-scale characteristics of the Galaxy. Kluwer, Dordrecht,
p. 451
%
\bibitem[1989]{ei:ha}
Einasto J., Haud U., 1989, A\&A 223, 89
%
\bibitem[1980]{einaa}
Einasto J., Tenjes P., Barabanov A.V., Zasov A.V., 1980,
Ap\&SS 67, 31
%
\bibitem[1998]{eise}
Eisenhauer F., Quirrenbach A., Zinnecker H., Grenzel R.,
1998, ApJ 498, 278
%
\bibitem[2000]{elme}
Elmegreen B.G., 2000, ApJ 530, 277
%
\bibitem[1997]{fe:wh}
Feast M., Whitelock P., 1997, MNRAS 291, 683
%
\bibitem[1990]{fi:tr}
Fish M., Tremaine S., 1990, ARA\&A 29, 409
%
\bibitem[1976]{ge:ge}
Georgelin Y.M, Georgelin Y.P, 1976, A\&A 49, 57
%
\bibitem[1988]{grab}
Grabelsky D.A., Cohen R.S., Bronfman L., Thaddeus P., 1988, ApJ 331,
181
%
\bibitem[1995]{hamb}
Hambly N.C., Fitzsimmons A., Keenan F.P., et al., 1995, ApJ
448, 628
%
\bibitem[1997]{hart}
Hartman J.W., 1997, A\&A 322, 127
%
\bibitem[1989]{ha:ei}
Haud U., Einasto J., 1989, A\&A 223, 59
%
\bibitem[2000]{hoog}
Hoogerwerf R., de Bruijne J.H.J., de Zeeuw P.T., 2000,
A\&A (in press) = astro-ph/0010057
%
\bibitem[1997]{kape}
Kaper L., van Loon J.Th., Augusteijn T., et al., 1997, ApJL 475,
L37
%
\bibitem[1986]{ke:ly}
Kerr F.J., Lynden-Bell D., 1986, MNRAS 221, 1023
%
\bibitem[2000]{krou}
Kroupa P., 2000, New Astron 4, 615
%
\bibitem[1956]{kuzm}
Kuzmin G., 1956, Astr. Zh. 33, 27
%
\bibitem[1991]{leon}
Leonard P.J.T., 1991, AJ 101, 562
%
\bibitem[1988]{le:du1}
Leonard P.J.T., Duncan M.J., 1988, AJ 96, 222
%
\bibitem[1990]{le:du2}
Leonard P.J.T., Duncan M.J., 1990, AJ 99, 608
%
\bibitem[1997]{lind}
Lindblad P.O., Lod\'en K., Palou\v{s} J., Lindegren L.,
1997. Runaway stars and the force perpendicular to the galactic
plane. In: Hipparcos -- Venice '97, ESA, p.~665 (=
//astro.estec.esa.nl/Hipparchos/venice.html)
%
\bibitem[1994]{ly:lo}
Lyne A.G., Lorimer D.R., 1994, Nature 369, 127
%
\bibitem[1998]{mait} 
Maitzen H.M., Paunzen E., Pressberger R.,
Slettebak A., Wagner R.M., 1998, A.\&A. 339, 782
%
\bibitem[1983]{mikk}
Mikkola S., 1983, MNRAS 203, 1107
%
\bibitem[1998]{moff}
Moffat A.F.J., Marchenko S.V., Seggewiss W., et al., 1998,
A.\&A. 331, 949
%
\bibitem[1999]{nele}
Nelemans G., Tauris T.M., van den Heuvel E.P.J., 1999, A\&A
352, L87
%
\bibitem[1995]{nezh}
Nezhinskij E.M., Ossipkov L.P., Kutuzov S.A., 1995. Open
cluster systems: kinematics, orbits. In: van der Kruit P.C.,
Gilmore G. (eds.), Stellar populations. Kluwer, Dordrecht, p.
374
%
\bibitem[1962]{ollo}
Ollongren A., 1962, BAN 16, 241
%
\bibitem[1954]{opik}
\"Opik E., 1954, Irish AJ 2, 219
%
\bibitem[1993]{ph:ja}
Phelps R.L., Janes K.A., 1993, AJ 106, 1870
%
\bibitem[1999]{port}
Portegies Zwart S.F., Makino J., McMillan S.L.W., Hut P.,
1999, A.\&A. 348, 117
%
\bibitem[1999]{pr:zi}
Preibisch T., Zinnecker H., 1999, AJ 117, 2381
%
\bibitem[1999]{roll}
Rolleston W.R.J., Hambly N.C., Keenan F.P., Dufton P.L.,
Saffer R.A., 1999, A.\&A. 347, 69
%
\bibitem[1999]{ryan}
Ryans R.S.I., Keenan F.P., Rolleston W.R.J., Sembach K.R.,
Davies R.D., 1999, MNRAS 304, 947
%
\bibitem[1997]{saff} 
Saffer R.A., Keenan F.P., Hambly N.C.,
Dufton P.L., Liebert J., 1997, ApJ 491, 172
%
\bibitem[2000]{smok}
Smoker J.V., Keenan F.P., Marcha M.J., Watson D., Irwin
M.J., 2000, A\&A 361, 60
%
\bibitem[1996]{ta:ba}
Tauris T.M., Bailes M., 1996, A\&A 315, 432
%
\bibitem[1998]{ta:ta}
Tauris T.M., Takens R.J., 1998, A\&A 330, 1047
%
\bibitem[1994]{tenj4}
Tenjes P., Haud U., Einasto J., 1994, A\&A 286, 753
%
\bibitem[1996]{tout}
Tout C.A., Pols O.R., Eggleton P.P., Han Z., 1996, MNRAS
281, 257
%
\bibitem[2000]{will}
Williams J.P., Blitz L., McKee C.F., 2000, The structure
and evolution of molecular clouds. In: Mannings V., Boss A.
(eds.) Protostars and planets IV. Tucson, Univ. Arizona Press
%
\end{thebibliography}
\end{document}